\documentclass[twocolumn,amssymb,aps,prl,shortbibliography]{revtex4-2}
\usepackage{amsfonts}
\usepackage[utf8]{inputenc}
\usepackage[english]{babel}
\usepackage{amsmath}
\usepackage{amsfonts}
\usepackage{bbm}
\usepackage{amssymb}
\usepackage{enumerate}
\usepackage{times,txfonts}
\usepackage{graphicx}
\usepackage{color}
\usepackage{amsmath}
\usepackage{bm}
\usepackage{xcolor}
\usepackage{array}
\usepackage{multirow}
\usepackage[colorlinks=false, linkbordercolor=red, citebordercolor=green, urlbordercolor=cyan, pdfborderstyle={/S/U/W 1}]{hyperref}
\usepackage[normalem]{ulem}
\usepackage{comment}

\begin{document}


\title{Interference of photons from independent hot atoms}

\author{Jarom\'{i}r Mika$^1$, Stuti Joshi$^1$, Luk\'{a}\v{s} Lachman$^1$, Robin Kaiser$^2$, Luk\'{a}\v{s} Slodi\v{c}ka$^{1}$}
\affiliation{$^1$ Department of Optics, Palack\'{y} University, 17. listopadu 12, 771 46 Olomouc, Czech Republic \\
$^2$ Universit\'e C\^ote d'Azur, CNRS, INPHYNI, 17 Rue Julien Laupr\^ete, 06200 Nice, France}



\begin{abstract}
The coherence of light from independent ensembles of elementary atomic emitters plays a paramount role in diverse areas of modern optics. We demonstrate the interference of photons scattered from independent ensembles of warm atoms in atomic vapor. It relies on the feasibility of the preservation of coherence of light scattered elastically in the forward and backward directions from Doppler-broadened atomic ensembles, such that photons with chaotic photon statistics from two opposite atomic velocity groups contribute to the same detection mode. While the random phase fluctuations of the scattered light caused by a large thermal motion prevent direct observability of the interference in the detected photon rate, the stable frequency difference between photons collected from scattering off counter-propagating laser beams provides strong periodic modulation of the photon coincidence rate with the period given by the detuning of the excitation laser from the atomic resonance. The presented interferometry represents a sensitive and robust methodology for Doppler-free optical atomic and molecular spectroscopy based on photon correlation measurements on scattered light.
\end{abstract}

\maketitle



\section{Introduction}
Optical coherence is paramount to all areas of modern optics. The feasibility of the coherence of light scattered from atoms enabled a broad range of essential applications ranging from optical atomic and molecular spectroscopy to diverse atomic sensors, efficient light-matter interfaces, and long-distance quantum communication~\cite{svanberg2001atomic,duan2001long,lukin2003colloquium,sangouard2011quantum,degen2017quantum,chaneliere2018quantum}. The corresponding experimental approaches and their theoretical analysis consistently acknowledge severely detrimental impact of thermal atomic motion as a predominant source of decoherence~\cite{carmichael1997coherence,ficek2005quantum,chou2005measurement,maunz2007quantum,eschner2001light,legero2004quantum,qian2016temporal,fedoseev2024coherent}.
The usual approaches towards mitigation of its impact correspond to suppression of thermal kinetic energy by means of laser cooling, the employment of tight atomic trapping potentials, optimized excitation geometries, or diverse spatial and spectral filtering schemes for minimization of the which-way information due to atomic recoils.

We propose and demonstrate the availability of preservation of optical coherence in the high-temperature limit of atomic scatterers by developing a scheme for generation and observation of the interference of light scattered from independent ensembles of atoms at room temperature. The observability of interference fringes is enabled by a broad spectral distribution of thermal atoms at high temperatures and employs a natural atomic velocity selectivity of photon scattering in a weak excitation regime~\cite{hughes2018velocity}. Excitation using a single laser beam in the retro-reflected spatial configuration allows for indistinguishable generation and observation of photons at two well-defined frequencies separated by the stable relative difference determined by the excitation laser detuning $\Delta_\mathrm{L}$ from the atomic resonance.
Due to the low phase-space density of the generated single-photon level light emerging from fast and inherently random processes related to the thermal atomic motion, the direct observation of the corresponding first-order interference is not feasible~\cite{mandel1965coherence,fano1961quantum,jordan1964quantum}. However, photons produced from such intrinsically independent sources can manifest the presence of transient interference visible in the second-order correlation function~\cite{mandel1964quantum,forrester1955photoelectric,magyar1963interference,lipsett1963coherence,pfleegor1967interference,zhai2006two,liu2014second,liu2016second,zhang2019photon,kim2020hong,romanova2022time,chen2010observation}.

We demonstrate the coherence of light scattering in hot atomic vapors within a broad range of detunings from atomic resonance with a positive contribution of different atomic velocity classes. The coherence can be maintained both in the forward and backward scattering, where the latter provides a stable frequency difference between the scattered fields and the employed atomic transition. The verification of coherence using the interferometric scheme including light fields scattered from counter-propagating excitation laser beams reveals intrinsic application in atomic spectroscopy. The methodology is based on the measurement of atomic fluorescence in photon correlation signal and allows for a sub-Doppler spectroscopy in low optical depth regime applicable to dilute ensembles of hot atoms. It allows for a direct estimation of the absolute value of the laser detuning from atomic resonance without any need for the modulation of the laser probe frequency or atomic transition parameters, and neither requires calibration of any interaction parameters or thermal atomic motion.


\section{Interferometry with hot atomic scatterers}
\begin{figure}[!t]
\centering\includegraphics[width=1\linewidth]{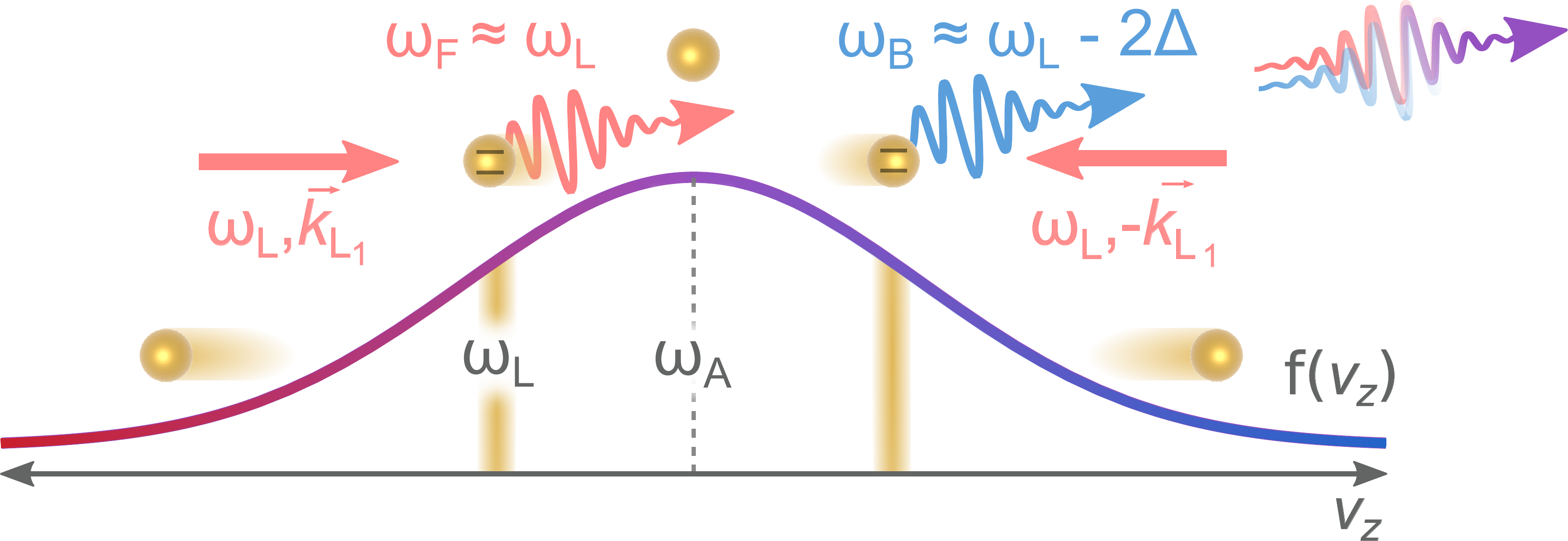}
\caption{The principle of interference of light from independent warm atomic ensembles. The laser at the frequency $\omega_\mathrm{L}$ is detuned by~$\Delta$ from the atomic transition $\omega_\mathrm{A}$ and scatters off the particular velocity class of atoms which follow the thermal velocity distribution $f(v)$ along the laser propagation axis. In the forward scattering~(F), the corresponding Doppler shift effectively compensates.
The retroreflected laser scatters off atoms possessing the opposite direction of motion but with the same velocity magnitude. The corresponding backward-scattered photons (B) are frequency-shifted by $\approx \Delta$ concerning an observer in the frame moving with the atomic scatterer and by approximately $2\Delta$ to the forward-scattered photons.}
\label{fig.main}
\end{figure}

The principle of the interferometric scheme is depicted in Fig.~\ref{fig.main}. The excitation laser beam with a wavevector $\textbf{k}_\mathrm{L_1}$ and frequency $\omega_\mathrm{L}$ is detuned from the atomic transition~$\omega_\mathrm{ A}$ by $\Delta$ and elastically scatters from hot thermal atoms with the particular projection of atomic velocity to the direction of the excitation laser $\textbf{v}_z$, which fulfills $\textbf{k}_\mathrm{ L_1}\cdot\textbf{v}\approx\Delta$.

The photons with a wave vector $\textbf{k}_\mathrm{F}$ scattered in close to a forward~(F) direction at an angle $\theta \to 0$ retain the central frequency of the laser $\omega_\mathrm{F} = \omega_\mathrm{L} +(\textbf{k}_\mathrm{L_1}-\textbf{k}_\mathrm{F})\cdot\textbf{v} \approx\omega_\mathrm{L}$. The standard deviation of their Gaussian spectral distribution corresponds to $\sigma_\mathrm{ F}=\sigma_\mathrm{ \omega,D} \sin\theta$, where $\sigma_\mathrm{\omega,D}$ is the width of the complete frequency redistributed spectrum for photon scattered in warm atomic vapor~\cite{hughes2018velocity,dussaux2016temporal,zhan2025room}. The scattered light is expected to present chaotic statistics with nearly ideal photon bunching $g^{(2)}(0)\approx 2$ in the single-mode detection regime~\cite{morisse2024temporal}. The second interferometer path is implemented by scattering the retro-reflected excitation laser with $\textbf{k}_\mathrm{L_2}=-\textbf{k}_\mathrm{L_1}$, which interacts with atoms with exactly opposite velocity direction~$-\textbf{v}_z$. Photons scattered backward~(B) with a wave vector $\textbf{k}_\mathrm{B}$ at an angle $\theta \to \pi$ towards the same detector will acquire the frequency $\omega_\mathrm{B} = \omega_\mathrm{L} + (\textbf{k}_\mathrm{L_2}-\textbf{k}_\mathrm{B})\cdot\textbf{v}$, which means, a frequency difference of $(\textbf{k}_\mathrm{L_2}-\textbf{k}_\mathrm{B})\cdot\textbf{v}\approx 2\Delta$ to the forward scattered part. Their spectral width results from the residual Doppler broadening for scattering at the angle $\theta+\pi$ and an additional term stemming from the longitudinal part of the velocity distribution $f(v_z)$ of contributing atoms.
In a low saturation limit, it corresponds to the natural linewidth~$\Gamma$ of the atomic transition. The total spectral width of back-scattered light then approximately results in $\sigma_\mathrm{B} \approx \sqrt{(\sigma_\mathrm{\omega,D} \sin\theta)^2+(2 \Gamma)^2}$, where the factor of two comes from the sum of the two equal Doppler shifts in the excitation and emission processes~\cite{hughes2018velocity}.
The fluctuating positions of atomic scatterers with Brownian motion prevent direct observability of interference of weak photon signals scattered from independent atoms in the detected photon rate. However, the stable frequency difference between forward and backward scattered fields can provide visible interference in the second-order correlations $g^{(2)}(\tau)$ with the characteristic beating with a period of $1/(2\Delta)$~\cite{mandel1964quantum}. The oscillations of photon correlations become observable for any detunings~$\Delta$ in the range $\sigma_\mathrm{\omega,D}> \Delta > \sigma_\mathrm{F}, \sigma_\mathrm{B}$. The upper limit is set by the width of the Doppler broadened atomic spectra $\sigma_\mathrm{\omega,D}$. The lower is given by the requirement on observation of modulation within the coherence time given by the residual Doppler broadening corresponding to~$\sigma_\mathrm{F}$ and $\sigma_\mathrm{B}$ or frequency difference to other electronic levels in hyperfine manifolds. Note that in addition to these fundamental limitations, particular technological constraints, including especially the temporal resolution of the employed photon detectors, or the overall detection efficiency, will set the upper limit on observable beating frequency~$2 \Delta$.

The single-mode optical field resulting from scattering on the two atomic velocity classes corresponding to the forward~(F) and backward~(B) -scattered light can be expressed as
\begin{equation}
E(t)\propto \sum_{i=1}^{N} E_{\mathrm{F}, i} e^{-\mathrm{i}(\omega_{\mathrm{F},i}t-\textbf{k}_{\mathrm{F}}\cdot\textbf{r}_i)} +\sum_{j=1}^{N} E_{\mathrm{B}, j} e^{-\mathrm{i}(\omega_{\mathrm{B},j}t-\textbf{k}_\mathrm{B}\cdot\textbf{r}_j)} .
\label{eq.field}
\end{equation}
where $E_{{\rm F}, i ({\rm B}, j)}$ are the amplitudes of the fields from $i (j)$-th atom, $\mathbf{r}$ are the position vectors of atoms. 
The summation includes contributions from a large number of~$N\gg1$ atoms in each of the two opposite atomic velocity classes. Here, we assume that the numbers of contributing atoms to the forward and backward scattered fields are approximately the same with relevance to the presented experimental configuration, where the numbers of contributing atoms are given by the observation volume of the two atomic ensembles with the same atomic density. The Doppler distribution of atomic velocity classes of thermal vapor affects the number of contributing atoms in both ensembles equally, as the effective laser detuning from atomic resonance is the same for both scattering contributions. The evaluation of the normalized second-order correlation function in the limit of thermally fluctuating atomic positions gives
\begin{equation}
g^{(2)}(\tau)=1+\frac{1}{(\bar{n}_\mathrm{ F}+\bar{n}_\mathrm{ B})^2}|\bar{n}_\mathrm{ F} g^{(1)}_\mathrm{ F}(\tau)+\bar{n}_\mathrm{ B} g^{(1)}_\mathrm{ B}(\tau)e^{\mathrm{i}(\omega_\mathrm{ F}-\omega_\mathrm{ B})\tau}|^2.
\label{eq.g2}
\end{equation}
Here, the first-order temporal correlation functions $g^{(1)}_\mathrm{F,B}(\tau)$ reflect the coherence of light scattered on the individual atomic ensembles. The term $\bar{n}_\mathrm{F(B)}$ correspond to mean photon number from a particular atomic ensemble, which can be affected by different atomic level populations and losses experienced by the two fields.
The complete derivation can be found in Supplementary Materials ~\cite{Supplement}, which includes references ~\cite{dussaux2016temporal,loudon2000quantum}. The equation reveals the crucial contribution of beating of the first-order correlation functions $g^{(1)}(\tau)$, which modulates the second-order photon correlations at their mean frequency difference $\omega_\mathrm{F}-\omega_\mathrm{B}$. It represents an extended version of the Siegert relation for chaotic light with two contributing atomic thermal light sources at different frequencies~\cite{loudon2000quantum}.
The visibility of the beating of first-order coherences can be controlled by the relative contributions of two atomic velocity classes, which depends on the amount of detected forward and backward scattered light. For equal number of photons from the two atomic velocity classes contributing to the signal, the contrast is maximal.
We emphasize that the resonant nature of the interaction is essential for the velocity selectivity of scatterers, as assumed in Eq.~(\ref{eq.field}). Consequently, high visibility of interference requires $(\omega_\mathrm{ F}-\omega_\mathrm{ B}) \approx 2 \Delta > \sigma_\mathrm{ F}, \sigma_\mathrm{ B}$, such that the emitted photons correspond to different non-overlapping spectral components with width given by the residual Doppler-broadening.

\section{Experimental demonstration}
\begin{figure}[!t]
\centering\includegraphics[width=1\linewidth]{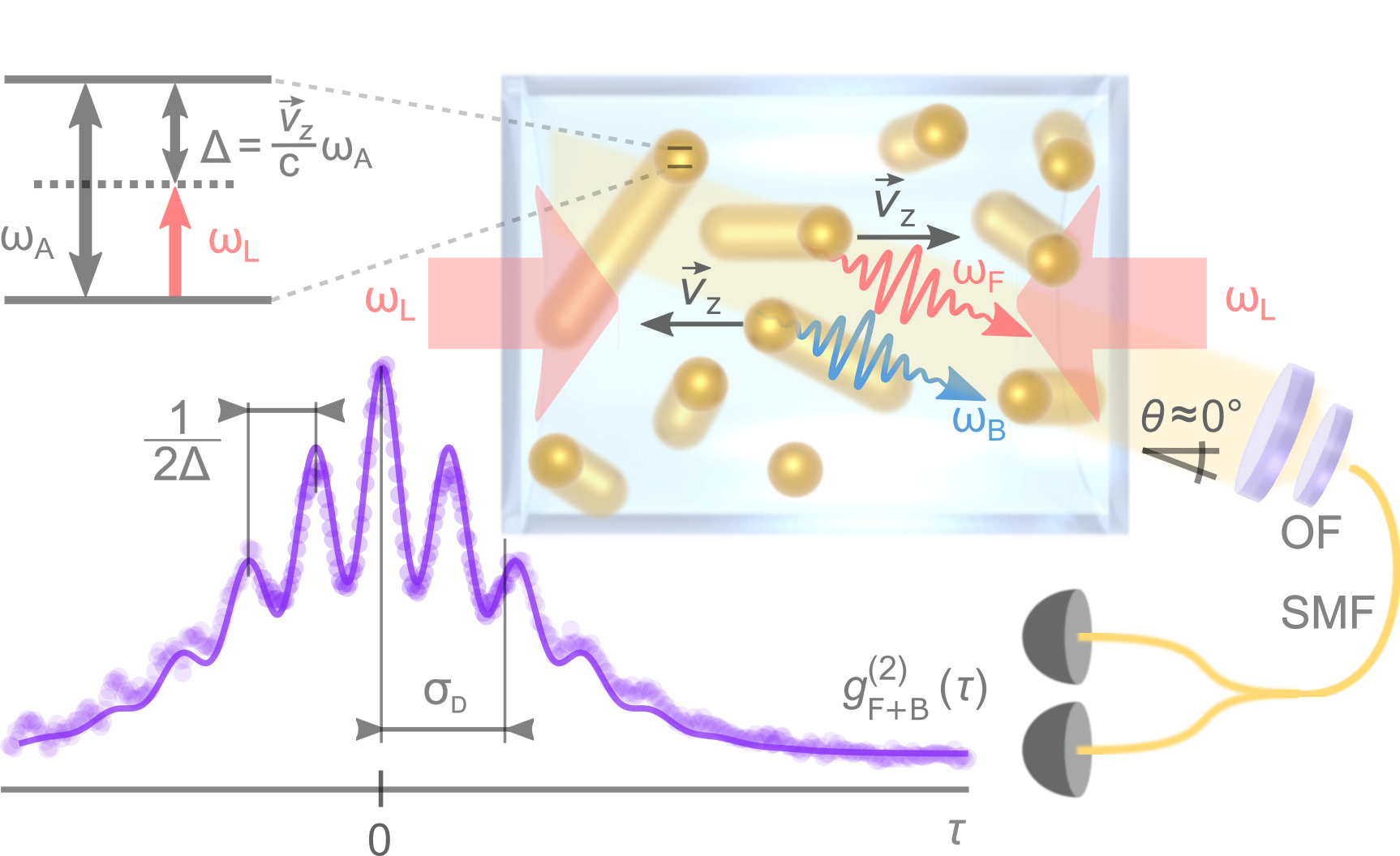}
\caption{Experimental scheme and example measurement of interference of light from warm atomic vapors. The scattering of two counter-propagating laser beams off the particular velocity classes of atoms results in forward~(F) and backward~(B) scattered photons with a frequency difference of $\approx 2\Delta$ collected in the same optical mode defined by the single mode optical fiber~(SMF). The measurement of the second-order correlations $g^{(2)}(\tau)$ provides coherent frequency beating with a period of $1/(2\Delta)$. The additional optical filtering (OF) in the detection path suppresses various noise contributions. Its detailed specification can be found in Supplementary materials ~\cite{Supplement}.}
\label{fig.main2}
\end{figure}
The experimental demonstration of the scheme for observation of interference from independent warm atoms employs the excitation of warm $^{87}$Rb vapor using a single laser beam in a retro-reflected standing-wave configuration, as displayed in Fig.~\ref{fig.main2}. A collimated Gaussian laser beam has a radius of $w_\mathrm{ E}=1.1 \pm 0.1$~mm and frequency detuning~$\Delta_\mathrm{ L}$ from the $5\mathrm{S}_{1/2}(F=2) \leftrightarrow 5\mathrm{P}_{3/2}(F'=3)$ transition. The optical power of the transmitted laser of $50\,\mu$W measured at the output of the cell along the forward-scattering direction corresponds to the excitation in a regime far from saturation. The standing wave configuration is achieved by the reflection from the planar mirror behind the cell.
The excitation and observation optical modes overlap in the proximity of the cell output viewport, which reduces photon losses for both interfering channels.
The forward-scattered photons are observed under the small angle $\theta = 2^\circ \pm 0.5^\circ$. The backward-scattered photons are collected into the same spatial mode secured by coupling to a single-mode optical fiber. The observation spatial Gaussian mode waist of~$w_0=95 \pm 5~\mu$m is positioned at the intersection with the excitation beam. The collected photons pass the polarization filter optimized for maximal transmission of the same circular polarization as the forward propagating excitation laser beam. We note that the selection of this particular polarization mode is not unique to the functioning of the scheme and it has been successfully tested also for linear polarization mode with similar results. Fabry-Pérot frequency filter is set up to suppress any residual light originating from Raman scatterings to the $F=1$ state manifold. A balanced set of two single-photon counting modules (SPCM) in a Hanbury-Brown-Twiss arrangement is employed for the analysis of the second-order photon correlations. The detection setup is assembled such that it provides the feasibility of simultaneous measurement of the modulus of the first-order coherence $|g^{(1)}(\tau)|$ in a Michelson interferometer, see Supplementary Materials ~\cite{Supplement}.

\begin{figure}[!t]
\centering\includegraphics[width=1\linewidth]{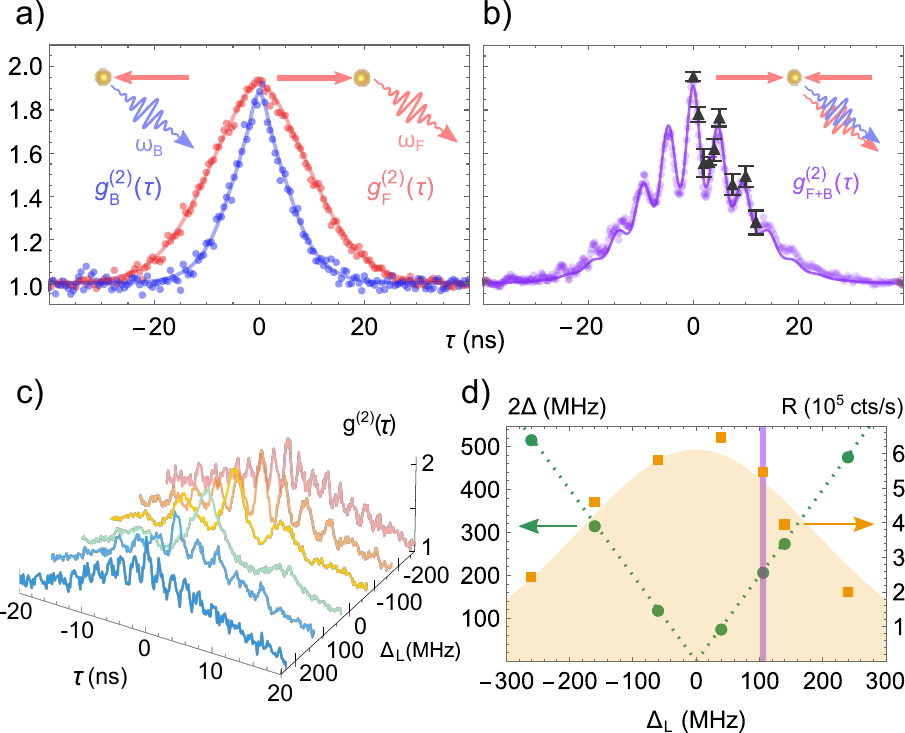}
\caption{a)~shows the measurements of the individual $g^{(2)}_\mathrm{ F(B)}(\tau)$ for photons emitted from two different velocity classes of ensembles of warm atoms. The corresponding $g^{(2)}_\mathrm{ F+B}(\tau)$ emerging from their interference is shown in~b). The fit of the interference pattern uses Eq.~(\ref{eq.g2}), where $\bar{g}^{(1)}(\tau)$ of atoms from the two atomic ensembles are taken from their independent measurements. The evaluation of particular $g^{(2)}(\tau)$ data points using the Siegert relation from the independently measured $|g^{(1)}(\tau)|^2+1$ is shown with black triangles. The error bars depict a single standard deviation. Graph in c)~summarizes measurements of $g^{(2)}(\tau)$ for different laser detunings $\Delta_\mathrm{L}$. The evaluated mean modulation frequencies $f_\mathrm{ mod}$ from a direct Fourier analysis are shown in~d). Here, green dotted lines illustrate its linear dependence on independently measured detuning~$\Delta_\mathrm{ L}$. The stability of such frequency estimation is practically limited by detected photon rates shown as orange data points. The filled area depicts the corresponding simulation considering the atomic populations at different velocity classes with a Gaussian uncertainty of the Doppler broadened spectra~$\sigma_\mathrm{ DB}$, including also the residual modification due to the Fabry-Pérot spectral filter. The marked data point corresponds to the measurement presented in b).}
\label{fig.results}
\end{figure}

Examples of the measured interference of light from independent warm atomic ensembles are shown in Fig.~\ref{fig.results}. The independent photon correlation measurements of forward and backward scattered fields for laser detuning $\Delta_\mathrm{ L}\approx 100$~MHz in~a) demonstrate a close-to ideal bunching values $g^{(2)}_\mathrm{ F}(0) = 1.94 \pm 0.02$ and $g^{(2)}_\mathrm{ B}(0) = 1.92 \pm 0.03$, emerging from the random phase shifts of scattered photons caused by the large thermal motion of atoms~\cite{loudon2000quantum,mika2018,morisse2024temporal}. Importantly, these individual correlation functions demonstrate large temporal coherences corresponding to spectral bandwidths $\sigma_\mathrm{ F} \approx 9.6$~MHz and $\sigma_\mathrm{ B}\approx 15.8$~MHz, respectively. They effectively correspond to average rates of random phase changes in each of the atomic ensembles and set an upper limit on the observable period of interference modulation in the corresponding~$g^{(2)}_\mathrm{ F+B}(\tau)$. The spectral widths agree with their independent evaluations within the uncertainty of measurement of the scattering angle~$\theta$ and confirm the predicted scaling of temporal coherence for the backward-scattered signal. We note that the presented backscattering measurements correspond to a first experimental study of $g^{(2)}_\mathrm{ B}(\tau)$ from warm atoms in such configuration and provide an important confirmation of the feasibility of coherence of light in this direction.
The plot in~b) depicts the corresponding $g^{(2)}_\mathrm{ F+B}(\tau)$, where both forward and backward-scattered fields contribute simultaneously. The interference results in clear modulation of photon coincidence detection probabilities for $|\tau|>0$ with the period of $f_\mathrm{ mod} = 210.8 \pm 1.2$~MHz, in agreement with the independently set laser detuning from the $F=2 \leftrightarrow F^{\prime}=3$. The periodic modulation between constructive and destructive interference at the frequency difference of interfering photons signifies the fulfillment of the critical conditions for the presented scheme. It confirms the sufficiency of the velocity selectivity in the given excitation regime and, at the same time, the mutually coherent contribution with sufficient coherence time from atoms at different velocity classes in a strongly Doppler-broadened thermal atomic ensemble.
The detected mean photon rate for presented excitation conditions was $R_\mathrm{(F+B)}=(64 \pm 2)\times 10^4$~counts/s.  The measured photon bunching at zero time delay $g^{(2)}_\mathrm{F+B}(0) = 1.96 \pm 0.01$. It allows for evaluating an upper limit on the ratio $r=0.02 \pm 0.005$ between the mean photon number of the residual phase-correlated light and the light from the two independent chaotic light sources with mutually randomized phases. The derivation of the simple model for quantifying residual phase correlations can be found in the Supplementary Information. ~\cite{Supplement} The rate of simultaneous detection of two photons $R_\mathrm{ C}=(4.9 \pm 0.3)\times 10^3$~coincidences/s for the time bin of 14.3~ns.

A complementary perspective on the emergence of the modulation of photon bunching can be obtained from the measurement of a first-order correlation function. The corresponding beating can still be observed in the $|g^{(1)}(\tau)|$ evaluated from the measured visibilities of the field autocorrelation measurement. However, rather than emerging from a direct phase interference between the photons emitted from the two disparate atomic velocity classes, the oscillations in $|g^{(1)}(\tau)|$ result from a periodic rephasing of the self-interference of these two independent narrowband fields for different relative time delays in the interferometer. The $g^{(2)}(\tau)$ evaluated from the measured $|g^{(1)}(\tau)|$ using a Siegert relation~\cite{eloy2018diffusing,loudon2000quantum} shown as black triangle data points in Fig.~\ref{fig.results} is in agreement with the independently measured second-order correlations within statistically estimated measurement uncertainties.

The fit of the data in Fig.~\ref{fig.results}-b) using the model based on Eq.~(\ref{eq.g2}) confirms, that the interference visibility of $32\pm 1$~\% is indeed determined by the relative difference in probabilities of detection of the forward to backward scattered photons. The corresponding photon rates were estimated from fit to $R_\mathrm{F}=(54\,\pm\,2)\times 10^4$~counts/s and $R_\mathrm{B}=(10 \pm 0.8)\times 10^4$~counts/s. When comparing to more than two orders of magnitude faster random phase changes in both modes $\sigma_\mathrm{ F}, \sigma_\mathrm{ B}$, this further validates the regime of interference between phase randomized photons. The large difference in the photon rates corresponds to the smaller intensity of the back-scattering laser beam, which experiences additional attenuation in the optically dense atomic sample before reaching the observed interaction volume in the cell. 
We note, that the feasibility of the modification of contributing photon rates by changing the polarization settings of the excitation beams and observation mode was tested, which allowed for observation of a high-visibility interference. We also tested the experimental geometry, where the transversal position shift of the excitation beam on the scale much beyond the beam diameter preserves the observability of the interference, which provides complementary proof of the independence of the two contributing atomic ensembles.




The presented interferometry has intrinsic potential for diverse applications in spectroscopy. It can be readily applied for a direct precise estimation of an absolute value of the relative frequency detuning between the excitation laser and internal atomic transition. 
The photon-correlation detection allows for direct measurement of the absolute value of laser detuning $\Delta$ from the Fourier analysis of $g^{(2)}(\tau)$ with a sub-natural linewidth precision. The estimation should also include the corresponding residual systematic shift of the beating frequency $f_\mathrm{ mod}=|\omega_\mathrm{ F}-\omega_\mathrm{ B}| = |\omega_L + \Delta (1 - \cos\theta)) - (\omega_L+\Delta (1+\cos(\theta+\pi))| = 2 \Delta \cos \theta$. The experimental geometry presented here with $\theta=2^\circ \pm 0.5^\circ$ implies a relative shift in the estimated frequency difference on the order of~$10^{-4}$. 
Fig.~\ref{fig.results}-c) presents the series of measured $g^{(2)}(\tau)$ with emergent linear dependence of the modulation frequency $f_\mathrm{ mod}$ on the independently estimated laser detuning~$\Delta_\mathrm{ L}$. Here, $\Delta_L$ was measured on the optical wavemeter with measured short-term instability below $\sigma \approx 1$~MHz. The linear fit using $f_\mathrm{ mod}=|\alpha\Delta_\mathrm{ L}|$ provides $\alpha=1.995 \pm 0.003$, with no apparent deviation from linearity in the measured range within experimental error bars. 
The evaluations of larger statistical sets at this particular setting corresponding to an average count rate of $(640\pm30)\times10^3$~counts/s provide rapid enhancement of stability for short measurement timescales, in close agreement with the $1/\sqrt{T}$ scaling, where $T$ is the averaging period. The stability of $0.3\pm 0.1$~MHz has been reached for $T=45$~s. 
The sampling resolution of the Fourier spectra is given by the maximal evaluated $\tau_\mathrm{ max}= 3.2\,\mu$s and corresponds to 0.154~MHz. The actual feasible frequency resolution is determined by the inverse of the coherence time of contributing photons. In the time domain, the temporal resolution of frequency estimation is set by the acquisition time necessary for achieving desired measurement stability, which is practically limited by the detectable photon rates. It severely decreases for large $\Delta$ which relies on the scattering of atoms from marginal velocity classes. The technical upper limit on the measurement bandwidth is set by the temporal jitter of the employed SPCMs with the average of~$500 \pm 100$~ps.

\section{Conclusions}
The presented experimental methodology provides an exclusive tool for observation of the emergence of coherence of light from independent ensembles of elementary atomic scatterers. It is directly applicable in a broad range of fundamental interferometric tests and in applications in optical atomic and molecular spectroscopy~\cite{svanberg2001atomic,demtroder2015laser,ludlow2015optical,pizzey2022laser,jensen2019detection}. It can be implemented to different classical or quantum scatterers with feasible resonant interaction, independently of the availability of their motional cooling. Importantly, contrary to the broad range of spectroscopy methods based on a free-space laser absorption with sub-Doppler resolution applicable to warm atoms, the presented interference scheme allows for implementation with small samples of atomic or molecular vapors and with electronic transitions providing only very low resonant optical depth.
Despite operating in a small saturation parameter limit, the resonant character of interaction is indispensable for its functioning, as it provides the necessary selectivity of velocity classes of contributing scatterers. For atoms, the observation of interference can thus be considered as a unique signature of the quantized internal structure in the elastic scattering limit. 
Importantly, the interference is not compromised by the large thermal motion. On the contrary, it enables a large measurement bandwidth of atomic spectroscopy.
The combination of a high atomic velocity selectivity and photon correlation detection provides sub-Doppler and sub-natural linewidth
precision in the regime of large thermal motion with demonstrated feasibility of implementation at the single photon level. The high atomic temperature enhances the spectral bandwidth of the spectroscopy, which differentiates from other laser spectroscopy methods applicable to hot atoms~\cite{svanberg2001atomic}. While the temporal resolution and stability of frequency estimation could in the presented example benefit from the large number of Rb atoms in a vapor cell, the scheme can be particularly useful for analysis of extremely small and dilute atomic or molecular samples due to its intrinsic stability stemming from the observation of the interference signal in second-order correlations. In contrast to the schemes employing direct observation of the first-order interference, the presented methodology is not affected by slow phase drifts of the excitation laser and of interfering signals. These features promise direct applicability for precision optical spectroscopy of dipole transitions of atomic vapors including only about $10^5$~atoms in the observation mode, which has the potential to enhance laser spectroscopy of diverse rare atomic isotopes with applicable energy level structure, for example, $^{46}$Ca, with relevance to applications~\cite{garcia2016unexpectedly,solaro2020improved}.
Extensions of presented elementary examples to atoms or molecules with closely spaced electronic level manifolds can result in complex beating structures providing information about the contributing energy level structure.

\section{Acknowledgments}
S.~J. and J.~M. acknowledge the support of the Czech Science Foundation under the project GA21-13265X. L.~L. acknowledges the support from the
Czech Science Foundation (Grant No. 23-06015O). L.~S. is grateful for the national funding from the MEYS under the project CZ.02.01.01/00/22 008/0004649. L.~S. and R.~K. acknowledge funding from the QUANTERA ERA-NET cofund in quantum technologies implemented within the European Union’s Horizon 2020 Programme (project PACE-IN, 8C20004). L.~S. is grateful to Romain Bachelard for stimulating discussions.

\section{Data availability}
Original experimental data that support the findings of this article can be found at Ref. ~\cite{zenodo}.

\bibliographystyle{apsrev4-2}
\bibliography{paper}

\newpage
\newpage\clearpage
\newpage

\renewcommand{\theequation}{S\arabic{equation}}
\renewcommand{\thefigure}{S\arabic{figure}}
\renewcommand{\thesection}{S\arabic{section}}
\renewcommand{\theHequation}{Supplement.\theequation}
\renewcommand{\theHfigure}{Supplement.\thefigure}
\renewcommand{\bibnumfmt}[1]{[S#1]}
\renewcommand{\citenumfont}[1]{S#1}

\setcounter{equation}{0}
\setcounter{figure}{0}
\setcounter{table}{0}
\setcounter{section}{0}
\setcounter{page}{1} \makeatletter

\section*{Supplementary information: Interference of photons from independent hot atoms}

\subsection{Experimental setup}

\begin{figure*}[!t]
\centering\includegraphics[width=1.0\linewidth]{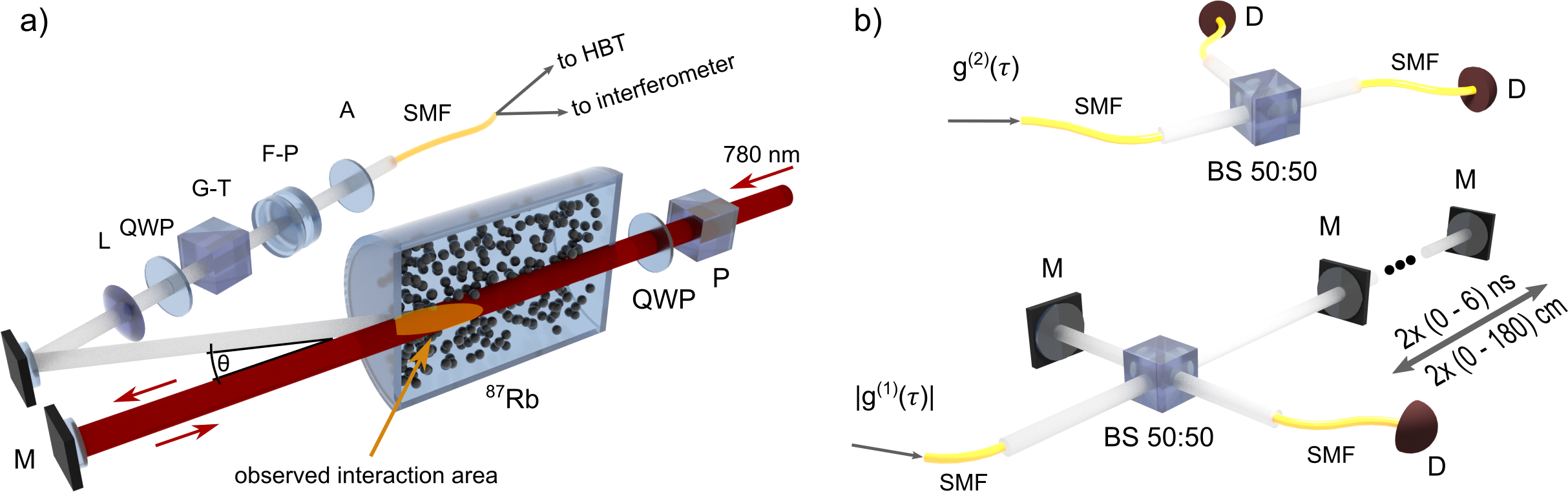}
\caption{Experimental scheme for observation of interference of light from warm atomic vapors. a) The excitation laser beam generated from an external-cavity diode laser at 780~nm is detuned by~$\Delta$ from the $5\mathrm{S}_{1/2}(F=2) \leftrightarrow 5\mathrm{P}_{3/2}(F'=3)$ transition of $^{87}$Rb. The laser alignment is set in a counter-propagating standing wave configuration. The scattered photons are observed in the single optical mode at angle~$\theta$ defined by the composite optical filter, including the spatial mode filter implemented by the combination of collecting lens~(L) and single-mode optical fiber~(SMF). The polarization mode coincident with the excitation laser polarization is set by the combination of the quarter waveplate~(QWP) and Glan-Thompson polarizer~(GT). The Fabry-Pérot~(F-P) resonator is set to suppress the contributions from residual Raman scattering. The optical attenuator~(A) corresponding to the optical neutral density filter was employed to suppress the detected count rate below the saturation limit of employed single photon detectors. The control of the position of the observed interaction area of lasers with atoms within the employed long atomic vapor cell allows for additional tunability and reduction of losses experienced by the scattered light upon traveling towards the detector. b) The collected photons are analyzed in the setup including two single-photon counting modules~(D) for measurement of the second-order correlations $g^{(2)}(\tau)$. Additional measurement of the degree of the modulus of the first-order coherence $|g^{(1)}(\tau)|$ is implemented in the Michelson interferometer with feasible relative arm length difference corresponding to up to $\tau=12$~ns.}
\label{setup}
\end{figure*}

The vapor of isotopically pure atoms is stored in a 7.5~cm long cylindrical glass cell with antireflection-coated input and output windows. The cell temperature is about $60^{\circ}$C. 
A collimated Gaussian laser beam with a circular polarization and a half width $w_{\rm E}=1.1 \pm 0.1$~mm scatters off atoms in a weak saturation limit. The frequency is set to detuning~$\Delta_{\rm L}$ from the $5\mathrm{S}_{1/2}(F=2) \leftrightarrow 5\mathrm{P}_{3/2}(F'=3)$ transition. The optical power is set to $50\,\mu$W measured at the output of the cell along the forward-scattering excitation beam. This corresponds to a small on-resonant saturation parameter $s\ll 1$. After the first pass of the vapor cell, the laser beam is retroreflected by a planar mirror, which results in an optical standing wave. As illustrated in Fig.~\ref{setup}, the excitation and observation optical modes overlap in the proximity of the cell output viewport, which reduces photon losses and the probability of multiple scattering~\cite{dussaux2016temporal}. The forward scattered photons are observed under the small angle $\theta = 2^\circ \pm 0.5^\circ$ and the backward scattered photons are collected into the exactly same spatial mode. The observation Gaussian spatial mode waist has a half width~$w_\mathrm{0}=95 \pm 5~\mu$m at the intersection with the excitation beam. The coupled mode is defined using a combination of a $f=300$~mm collection lens, which is employed for the collimation of coupled photons, followed by coupling to a single-mode optical fiber using an aspheric lens. The collected photons pass the polarization filter consisting of a quarter-wave plate and the Glan-Thompson polarizer optimized for the maximal transmission of the circular polarization with the same helicity as the forward propagating excitation laser beam. We note that the selection of this particular polarization mode is not unique to the functioning of the scheme in the presented experimental settings, and it has been successfully tested also for linear polarization mode with similar results. The Fabry-Pérot filter with a full width at a half maximum linewidth of $900 \pm 100$ MHz and the free spectral range of 30~GHz is set up to suppress any residual light originating from Raman scattering to the $F=1$ state manifold. The overall losses of photons scattered into the detection spatial mode were estimated to be about~26~\%, which includes the transmission of all optical components and the efficiency of the employed single-photon counting modules.

The collected light is decoupled from the single-mode fiber into a free-space Hanbury-Brown-Twiss detection arrangement employing a balanced coupling to two single-photon avalanche photodiodes for analysis of the second-order correlation functions. The implemented regime results in the detected mean photon rate of $(64 \pm 2)\times 10^4$~counts/s and the corresponding rate of simultaneous detection of two photons for the time bin of 14.3~ns is $4900\pm 300$~coincidences/s. The chosen time bin corresponds to the full width at the half maximum of the envelope of measured $g^{(2)}_{\rm F+B}(\tau)$. The detection setup is assembled such that it additionally provides the feasibility of simultaneous measurement of the modulus of the first-order coherence $|g^{(1)}(\tau)|$ in a free-space Michelson interferometer. We note, that the saturation limit of the employed single photon detectors necessitated the installation of an additional optical attenuation component based on polarization optics in the detection optical mode in the presented interaction parameters configuration. For the measurements presented here, its transmission efficiency factor was set to $\approx 0.25$. The separate measurements of $g^{(2)}_{\rm F}(\tau)$ presented in Fig.~3-a) in the main part of the manuscript were realized by simply blocking the back-reflected beam. The corresponding $g^{(2)}_{\rm B}(\tau)$ from solely back-reflected photon signal was achieved by implementing an additional separate beam with the same optical power and wavevector corresponding to exactly $\vec{k}_{L_2}$. Here, the same excitation spatial mode was guaranteed by coupling the excitation beam to the input single-mode fiber. The implementations of the excitation standing wave with a mirror, or, by employing a second excitation beam, are fully analogous. However, the employment of the mirror offers a simpler alignment procedure.

As the two ensembles of atoms are intrinsically independent, the contributing mean photon numbers $\bar{n}_{\rm F(B)}$ can be independently modified by controlling the respective excitation probabilities by adjusting the relative excitation powers of the corresponding laser beams. This can be simply achieved by spatially displacing the retro-reflected laser beam, or by changing its polarization. The excitation geometry, where the backward-scattering atomic ensemble is closer to the detector setup simultaneously places its overlap with the observation spatial mode closer to the vapor cell window.
Consequently, the back-scattered photons experience less absorption, which can be also employed for balancing the photon rates. We experimentally confirmed that the transversal beam position shift on the scale much beyond the beam diameter preserves the observability of the interference, which provides complementary proof of the intrinsic independence of the two contributing atomic ensembles. As a result of these optimizations, the intensity ratio of the forward and backward-scattered photons became close to equal and the interference visibility reached more than~90~\%.

\subsection{Second-order coherence from warm atoms}

We consider scattering of two light fields with wave vectors $\mathbf{k}_{\rm L_1}$ and $\mathbf{k}_{\rm L_2}=-\mathbf{k}_{\rm L_1}$ on an ensemble of hot atoms.
The light fields scattered off the atomic ensembles with two opposite velocity classes have observable frequencies $\omega_{{\rm F},i}=\omega_{\rm L} +(\mathbf{k}_{\rm L_1}-\mathbf{k}_{\rm F})\cdot\mathbf{v}$ and $\omega_{{\rm B},i}=\omega_{\rm L} + (\mathbf{k}_{\rm L_2}-\mathbf{k}_{\rm B})\cdot\mathbf{v}$, corresponding to the forward~(F) and backward~(B) scattered photons from $i$-th and $j$-th atoms, respectively, with wave vectors $\vec{k}_{\rm F}$ and $\mathbf{k}_{\rm B}$ of the forward and backward scattered light. Their coupling to the same single optical mode results in
\begin{equation}
E(t)\propto \sum_{i=1}^{N}E_{{\rm F},i} e^{-\mathrm{i}\left(\omega_{{\rm F},i}t-\phi_{F,i}\right)}+ \sum_{j=1}^{N}E_{{\rm B},j} e^{-\mathrm{i}\left(\omega_{{\rm B},j}t-\phi_{B,j}\right)}
\label{eq.fieldS}
\end{equation}
where $\phi_{F,i}=\left(\textbf{k}_{\rm F}-\textbf{k}_{\rm L_1}\right)\cdot\textbf{r}_i$, $\phi_{B,j}=\left(\textbf{k}_{\rm B}-\textbf{k}_{\rm L_2}\right)\cdot\textbf{r}_j$ characterize dependence of a phase on atomic position vectors $r_i$ ($r_j$), and $E_{{\rm F}, i ({\rm B}, j)}$ are the amplitudes of the fields from $i (j)$-th atom. The summation includes contributions from~$N$ atoms in each of the two velocity classes. With relevance to the presented experimental configuration, the number of contributing atoms to the forward and backward scattered fields is assumed to be equal. It is determined by the excitation-observation mode overlap of the two atomic ensembles, which is set mostly by the same diameter of the excitation laser beam and of the observation Gaussian mode. 
Due to coupling the scattered light through a small angle $\theta=2^{\circ}$, we approximate the phases $\phi_{F,i}$ and $\phi_{B,j}$ in Eq.~(\ref{eq.fieldS}) as $\phi_{F,i}\approx |\mathbf{k}_{L_1}||\mathbf{r}_i|\theta^2/2$ and $\phi_{B,j}\approx 2|\mathbf{k}_{L_1}||\mathbf{r}_j|$. Scattering on an atom with a random position within the interaction region implies random changes of both phases $\phi_{F,i}$ and $\phi_{B,j}$.
The evaluation of the second-order correlation function
\begin{equation}\label{SM.G2}
G^{(2)}(\tau)=\langle E^{\ast}(t)E^{\ast}(t+\tau)E(t+\tau)E(t) \rangle
\end{equation}
follows the expansion to different combinations of fields in Eq.~(\ref{eq.fieldS}). Specific combinations of these fields in the expansion yield members that depend on the phase $\Delta_{F,i,j}=\phi_{F,i}-\phi_{F,j}$ or $\Delta_{B,i,j}=\phi_{B,i}-\phi_{B,j}$ obeying $\Delta_{F,i,j}\in (0,|\mathbf{k}_{L_1}|d\theta^2/2)$ and $\Delta_{B,i,j}\in (0,|\mathbf{k}_{L_2}|d)$. The length of the interaction region $d=2$ cm leads to the experimental values $|\mathbf{k}_{L_1}|d\theta^2/2=31.2 \pi$ and $|\mathbf{k}_{L_2}|d=102.6\times 10^3 \pi$. Fast movement of warm atoms eliminate any correlation among atomic positions during consecutive detection events. Thus, the corresponding phase-dependent members in expansion of Eq.~(\ref{SM.G2}) average to zero because $\Delta_{F,i,j}$ and $\Delta_{B,i,j}$ fluctuate due to the atomic movement over a range  that significantly exceeds $2 \pi$. This simplifies the expansion of $G^{(2)}(\tau)$ following as:  
\begin{widetext}
\begin{eqnarray}
G^{(2)}(\tau) & = & \sum_{i}\langle E_{{\rm F}, i}^*(t)E_{{\rm F}, i}^*(t+\tau)E_{{\rm F}, i}(t+\tau)E_{{\rm F}, i}(t) \rangle
+\sum_{j}\langle E_{{\rm F}, j}^*(t)E_{{\rm F}, j}^*(t+\tau)E_{{\rm F}, j}(t+\tau)E_{{\rm F}, j}(t)\rangle\\ \nonumber
&+&\sum_{i\neq l}\big\{\langle E_{{\rm F}, i}^*(t)E_{{\rm F}, i}(t+\tau)\rangle\langle E_{{\rm F}, l}^*(t+\tau)E_{{\rm F}, l}(t)\rangle+\langle E_{{\rm F}, i}^*(t)E_{{\rm F}, i}(t)\rangle\langle E_{{\rm F}, l}^*(t+\tau)E_{{\rm F}, l}(t+\tau)\rangle\big\}\\ \nonumber
&+&\sum_{j\neq m}\big\{\langle E_{{\rm B}, j}^*(t)E_{{\rm B}, j}(t+\tau)\rangle\langle E_{{\rm B}, m}^*(t+\tau)E_{{\rm B}, m}(t)\rangle+\langle E_{{\rm B}, j}^*(t)E_{{\rm B}, j}(t)\rangle\langle E_{{\rm B}, m}^*(t+\tau)E_{{\rm B}, m}(t+\tau)\rangle\big\} \\ \nonumber
&+&\sum_{i,j}\big\{\langle E_{{\rm F}, i}^*(t)E_{{\rm F}, i}(t+\tau)\rangle\langle E_{{\rm B}, j}^*(t+\tau)E_{{\rm B}, j}(t)\rangle
+\langle E_{{\rm F}, i}^*(t+\tau)E_{{\rm F}, i}(t)\rangle\langle E_{{\rm B}, j}^*(t)E_{{\rm B}, j}(t+\tau)\rangle\\ \nonumber
&\,&\,\,\,\,\,\,\,\,\,\,\,+\langle E_{{\rm F}, i}^*(t)E_{{\rm F}, i}(t)\rangle\langle E_{{\rm B}, j}^*(t+\tau)E_{{\rm B}, j}(t+\tau)\rangle+
\langle E_{{\rm B}, j}^*(t)E_{{\rm B}, j}(t)\rangle\langle E_{{\rm F}, i}^*(t+\tau)E_{{\rm F}, i}(t+\tau)\rangle\big\},\\ \nonumber
\label{longG2}
\label{eq.S3}
\end{eqnarray}
\end{widetext}
where we ignored all members of the expansion that depend on $\Delta_{F,i,j}$ or $\Delta_{B,i,j}$.  
In the remaining correlation expressions, the multiplication of position-dependent phase factors from expression~(\ref{eq.fieldS}) resulted in unity factors. In the limit of a large number of contributing atoms $(N\gg 1)$, the distribution of atomic velocity classes and of corresponding observed photon frequencies $\omega_{\rm F (B)}$ become continuous
\begin{equation}
\sum\langle E^*(t)E(t+\tau)\rangle \rightarrow \int \langle E^*(t)E(t+\tau)\rangle d\omega.
\end{equation}
The resulting second-order correlation function can then be expressed as
\begin{eqnarray}
\label{g2largeS}
G^{(2)}(\tau)& = & N\big(G^{(2)}_{\rm F}(\tau) +G^{(2)}_{\rm B}(\tau)\big)  \\ \nonumber
& + & N^2|G_{\rm F}^{(1)}(\tau)+G_{\rm B}^{(1)}(\tau)e^{\mathrm{i}(\omega_{\rm F}-\omega_{\rm B})\tau}|^2 \\
& + & \big(\bar{n}_{\rm F}+\bar{n}_{\rm B}\big)^2,\nonumber
\end{eqnarray}
where, $G^{(1)}_{\rm F(B)}(\tau)$ and $G^{(2)}_{\rm F(B)}(\tau)$ are the first- and second-order correlation functions of contributing individual atomic scatterers, and $\bar{n}_{\rm F(B)}$ are the mean photon numbers from a particular ensemble.
The Eq.~(\ref{g2largeS}) reveals the crucial contribution of the beating of the first-order correlation functions, which modulates the second-order photon correlation at the relative mean frequency difference $\omega_{\rm F}-\omega_{\rm B}$. The corresponding normalized second-order correlation function reads
\begin{equation}
\label{g2S}
g^{(2)}(\tau)=1+\frac{1}{(\bar{n}_{\rm F}+\bar{n}_{\rm B})^2}|\bar{n}_{\rm F} g^{(1)}_{\rm F}(\tau)+\bar{n}_{\rm B} g^{(1)}_{\rm B}(\tau)e^{\mathrm{i}(\omega_{\rm F}-\omega_{\rm B})\tau}|^2,
\end{equation}
where $g^{(1)}(\tau)$ are the normalized first-order correlation functions of light. The Eq.~(\ref{g2S}) represents an extended version of the Siegert relation for chaotic light~\cite{loudon2000quantum} for contributing two atomic thermal light sources at frequencies $\omega_{\rm F}$ and $\omega_{\rm B}$.
We note that the same single-mode limit can be approached also by $\bar{n}_{\rm F (or\,B)}\approx~0$, which can be experimentally observed by, for example, frequency filtering of one of the contributing atomic groups or by blocking the corresponding excitation laser beam.

\subsection{Model of radiation with partial phase correlations}

We introduce a simple model describing the reduction of the second-order correlation function $g^{(2)}(0)$ by considering mixing of the light scattered from warm atoms with the phase-correlated light corresponding to the coherent state $|\alpha\rangle$. It allows for evaluation of a lower limit of the contribution of light from scatterers with mutually fully randomized phases of their optical field contributions to the detected mode. This phase randomization of excitation lasers secures the independence of the two interfering fields. As described in the main manuscript part, it emerges from different contributions of thermal atomic motion, manifesting in random atomic positions and Doppler shifts, which result in effective spectral widths $\sigma_\mathrm{F}, \sigma_\mathrm{B}$ of the interfering light fields. The introduction of the fully phase-correlated state - coherent state $|\alpha\rangle$ is characterized by its well-defined phase and effectively accounts for all residual contributions to the detected mode, other than the dominant phase randomized fields with chaotic statistics scattered in the forward and backward directions. These include the residual scattering of the excitation laser light on atomic vapor cell windows to the detected mode, residual temporal multi-modeness due to the finite temporal resolution of the employed single-photon counting modules, and their noise - dark counts.
The corresponding $g^{(2)}(0)$ can be expressed as
\begin{equation}\label{SM:g2Coh}
	g^{(2)}(0)=\frac{G^{(2)}(0)+2\bar{n} |\alpha|^2+|\alpha|^4}{\left(\bar{n} +|\alpha|^2\right)^2},
\end{equation}
where $G^{(2)}(0)=2 \bar{n}^2$ is given by Eq.~(\ref{g2largeS}),
$\bar{n}=(\bar{n}_F+\bar{n}_B)$ represents the mean photon number of the scattered light and $|\alpha|^2$ is a mean photon number of the coherent state. Let 
$r=|\alpha|^2/ \bar{n}$ denote the ratio of the mean photon numbers of phase-correlated, i.e. coherent light to the light scattered from atoms. Eq.~(\ref{SM:g2Coh}) can then be simplified to
\begin{equation}\label{SM:g2r}
	g^{(2)}(0)= \frac{2 + 2 r+r^2}{(1+r)^2}.
\end{equation}
Eq.~(\ref{SM:g2r}) accounts for the suppression of $g^{(2)}(0)$ using the single parameter - ratio $r$. For fully phase randomized contributions, $g^{(2)}(0)=2$ and $r=0$, while we obtain $g^{(2)}(0)=1$ in the limit $r\rightarrow \infty$ corresponding to observation of phase-correlated laser light.

\section{Detuning $\Delta_\mathrm{L}$}

\begin{figure}[!h]
\centering\includegraphics[width=1.0\linewidth]{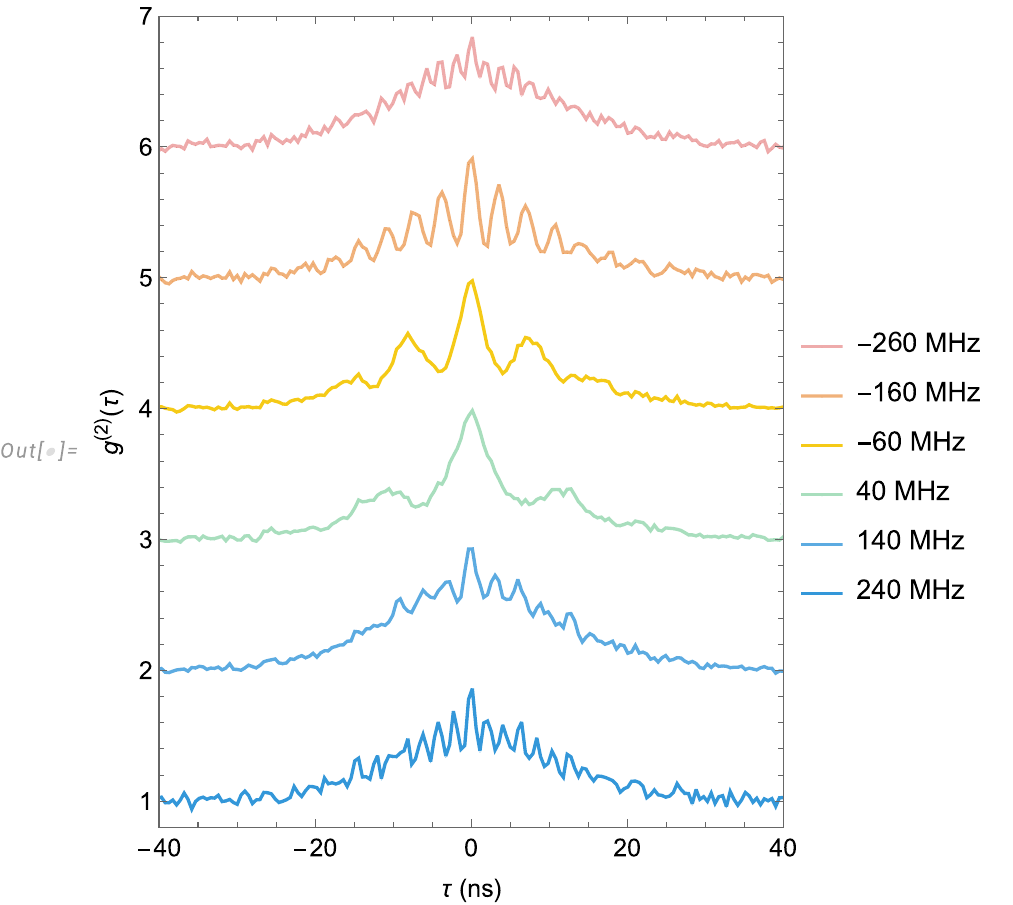}
\caption{Second order correlation functions $g^{(2)}(\tau)$ for various detunings $\Delta_\mathrm{L}$ of the laser from the employed transition. For better visualisation, the values of $g^{(2)}(\tau)$ are shifted. The error bars for the noisiest dataset (-40 MHz) reach $\pm0.03$ and the main noise contribution comes from shot noise.}
\label{fig:detuning}
\end{figure}

The feasibility of observing frequency beating in $g^{(2)}(\tau)$ relies on the detuning $\Delta_\mathrm{L}$ of the excitation laser from the involved atomic transition, which must fulfil the condition $\sigma_\mathrm{\omega,D}>\Delta_\mathrm{L}>\sigma_\mathrm{F},\sigma_\mathrm{B}$. In practice, the upper limit on the resolved beating frequency is also given by the jitter of the employed detectors and photon counting devices, which are of the same order as is the Doppler broadened atomic spectra $\sigma_\mathrm{\omega,D}\approx 228$ MHz for $60^{\circ}$C. Figure \ref{fig:detuning} shows the $g^{(2)}(\tau)$ for range of detunings $\Delta_\mathrm{L}$. The loss of contrast for higher absolute vaues of detuning is caused by both resolution limits of the employed electronics and by the change of detected forward-to-backward photon ratio.

\end{document}